\documentclass[twocolumn,showpacs,superscriptaddress,preprintnumbers,amsmath,amssymb,epsfig,floatfix,prx]{revtex4-1}

\usepackage{pdfpages}
\usepackage{pgffor}
\usepackage{etoolbox} 
\makeatletter
\patchcmd{\@outputpage@head}{\@ifx{\LS@rot\@undefined}{}{\LS@rot}}{}{}{}
\makeatother

\usepackage{amsmath}
\usepackage{multirow}
\usepackage{amsfonts}
\usepackage{amssymb}
\usepackage{calligra}
\usepackage{calrsfs}
\DeclareMathAlphabet{\mathcalligra}{T1}{calligra}{l}{m}
\usepackage{epsfig}
\usepackage{graphicx}
\usepackage{dcolumn}
\usepackage{color}
\usepackage{natbib}  
\usepackage{hyperref}
\usepackage{breakurl}
\hypersetup{colorlinks=true, citecolor=blue, urlcolor=blue, linkcolor=blue}
\usepackage{bm}
\usepackage{tabularx}
\newcolumntype{L}[1]{>{\raggedright\arraybackslash}p{#1}}
\newcolumntype{C}[1]{>{\centering\arraybackslash}p{#1}}
\newcolumntype{R}[1]{>{\raggedleft\arraybackslash}p{#1}}

\begin{document}
\title{Exciton in phosphorene: Strain, impurity, thickness and heterostructure}
\author{Srilatha Arra}
\thanks{These authors contributed equally}
\affiliation{Department of Chemistry, Indian Institute of Science Education and Research, Pune 411008, India}
\author{Rohit Babar}
\thanks{These authors contributed equally}
\affiliation{Department of Physics, Indian Institute of Science Education and Research, Pune 411008, India}\author{Mukul Kabir}
\email{Corresponding author: mukul.kabir@iiserpune.ac.in} 
\affiliation{Department of Physics, Indian Institute of Science Education and Research, Pune 411008, India}
\affiliation{Centre for Energy Science, Indian Institute of Science Education and Research, Pune 411008, India}
\date{\today}

\begin{abstract} 
Reduced electron screening in two-dimension plays a fundamental role in determining exciton properties, which dictates optoelectronic and photonic device performances. Considering the explicit electron-hole interaction within the $GW-$Bethe-Salpeter formalism, we first study the excitonic properties of pristine phosphorene and investigate the effects of strain and impurity coverage. The calculations reveal strongly bound exciton in these systems with anisotropic spatial delocalization. Further, we present a simplified hydrogenic model with anisotropic exciton mass and effective electron screening as parameters, and the corresponding results are in excellent agreement with the present $GW-$BSE calculations. The simplified model is then used to investigate exciton renormalization in few-layer and heterostructure phosphorene. The changes in carrier effective mass along with increasing electron screening renormalizes the exciton binding in these systems. We establish that the present model, where the parameters are calculated within computationally less expensive first-principles calculations, can predict exciton properties with excellent accuracy for larger two-dimensional systems, where the many-body $GW-$BSE calculations are impossible. 
\end{abstract}
\maketitle

\section{Introduction}
Excitonic properties of two-dimensional (2D) materials are markedly different from their bulk counterparts due to the fundamental difference in electron screening, and has attracted much attention conceptually in recent times.~\cite{PhysRevLett.115.066403,PhysRevLett.116.066803,PhysRevLett.116.056401} Many 2D materials interact strongly with light, and the concurrently generated electron-hole pairs interact strongly due to reduced screening.~\citep{nphoton.2015.282,nl0716404, nl501133c, nl504868p, srep06608, nmat4061} For example, while exciton binding is small, 84 meV, in bulk MoS$_2$,~\citep{APL1.4945047} due to reduced screening  in monolayer MoS$_2$, it is measured to be in 220--570 meV range.~\citep{nl501133c,nl504868p,srep06608}  Moreover, unlike the Frenkel excitons, the excitons in 2D materials can be delocalized in space and extend over 1 nm.~\citep{nl0716404, nmat4061,PhysRevLett.111.216805}  Further, the widely varying exciton binding has been reported in the different class of materials and has important implications in device applications. For applications in solar cells, photodetectors, and catalytic devices, exciton with weak binding leading to easy dissociation is desirable. In contrast, the materials with strong exciton binding are ideal to study plausible exciton-polariton condensate and for applications such as polariton lasing.~\citep{nphys3143}

Phosphorene has attracted a lot of attention due to its many plausible technological applications,~\citep{10.1038/nnano.2014.35, nn501226z, acs.jpcc.6b05069, natrevmats.2016.61,10.1021/acs.jpcc.7b12649} and has become an interesting proving ground for many-body physics.\citep{nl5043769, Kim723, PhysRevB.97.045132} The Dirac semimetal state has been experimentally realized in few-layer phosphorene under adatom absorption.~\citep{Kim723} We have recently reported an intrinsic, robust and high-temperature Kondo state in defected phosphorene doped with a transition-metal impurity.~\citep{PhysRevB.97.045132} In the present context, phosphorene has unique optical properties that are mainly determined by the quasiparticle band structure and screening. Originating from a puckered honeycomb network, the electronic band structure is highly asymmetric in phosphorene and consequently, the optical properties are also found to be highly anisotropic.~\citep{nn501226z,PhysRevLett.115.066403,acs.jpcc.6b05069,nnano.2015.71} The quasiparticle and optical gaps of single-layer phosphorene (SLP) are experimentally measured to be 2.2 $\pm$ 0.1 eV and 1.3 $\pm$ 0.02, respectively, through photo-luminescence excitation spectroscopy (PLE).~\citep{nnano.2015.71} This results in a very high exciton binding energy of 0.9 $\pm$ 0.12 eV due to reduced screening in 2D quantum confinement.  

Studying quasiparticle band structure and the corresponding optical properties is a non-trivial and computationally expensive task.  The conventional density functional approach fails to reproduce the correct experimental results which involve excited states.~\citep{RevModPhys.74.601}  In contrast, the many-body perturbation theory based $GW$ method produces the correct quasiparticle energies.~\citep{PhysRev.139.A796}   In this approach, the electron self-energies are expressed in terms of Green's function $G$ and screened Coulomb interaction $W$. Further, the optical properties of semiconductors and insulators are strongly affected by interacting electron-hole pairs, which is described through the Bethe-Salpeter equation (BSE).~\citep{PhysRev.84.1232, PhysRevLett.80.4510,PhysRevLett.81.2312} These theoretical descriptions lead to an excellent agreement with the experimental results.~\citep{RevModPhys.74.601}

Here, we study the quasiparticle and optical properties of SLP and its derivatives within the $GW$ and BSE formalisms, and also investigate the effect of strain.   Results on the pristine SLP are in excellent agreement with those obtained from the transmission and photoluminescence spectroscopies.~\citep{nl502892t,nnano.2015.71} The optical absorption and the corresponding excitons are found to be highly anisotropic. Due to the 2D confinement of photogenerated electron-hole pairs, the exciton binding is exceptionally strong, which is in accordance with the experimental measurements.~\citep{nnano.2015.71}  

While a rigorous treatment of electron-hole interaction within the BSE formalism provides an excellent description of exciton binding,~\citep{RevModPhys.74.601,PhysRevLett.111.216805,nmat4061,PhysRevB.87.155304,PhysRevB.86.115409} it is computationally very expensive and thus restricted to the systems with small size.  In this context, we describe a hydrogenic effective exciton mass model in which the parameters, the effective carrier masses and the static dielectric constant, are calculated within the conventional density functional theory (DFT) calculations.~\citep{PhysRevB.91.245421,PhysRevB.95.235434} The results of this simplified model are in excellent agreement with those calculated within the BSE formalism for the pristine and strained SLP along with the SLPs with 1 ML impurity coverage. Further, the anisotropic hydrogenic model is extended for larger systems with low impurity coverages that are earlier predicted to be good candidate materials for water redox reactions.~\citep{10.1021/acs.jpcc.7b12649}

The quantum confinement of exciton should be extraordinarily affected by the varied thickness of the 2D material, which alters electron screening. This picture is well captured within the present model through the renormalization of exciton binding in few-layer phosphorene. Furthermore, the practical applications of phosphorene are limited due to its fast degradation in ambient conditions resulting in a severe alteration in the corresponding electronic properties.~\citep{nl5032293,2053-1583-2-1-011002,nmat4299} Thus to avoid degradation, phosphorene is encapsulated with a capping layer and substrate,  and the devices restore the intrinsic carrier mobility of phosphorene.~\citep{nnano.2015.71,nl5032293,10.1038/nnano.2014.35,acsnano.5b00289,ncomms7647,ncomms8315} In this regard, we also investigate the electronic and exciton properties in SLP encapsulated with atomically thin hexagonal boron-nitride (h-BN), which is often used to protect phosphorene from degradation.~\citep{acsnano.5b00289,ncomms7647,ncomms8315}

\section{Methodology}
The structural optimizations were carried out within the conventional DFT formalism~\citep{PhysRevB.47.558,PhysRevB.54.11169} and the electrons are treated within the projector augmented wave method.~\citep{PhysRevB.50.17953} The Kohn-Sham orbitals were expanded in plane-wave basis with 400 eV energy cutoff. The exchange-correlation energy was described with Perdew-Burke-Ernzerhof (PBE) functional.~\citep{PhysRevLett.77.3865} The Brillouin-zone was sampled using $\Gamma$-centred 17$\times$13$\times$1 Monkhorst-Pack $k$-grid.~\citep{PhysRevB.13.5188} Complete structural optimization was carried until the forces exerted on each atom are less than 0.01 eV/\AA\ threshold. In phosphorene, three electrons participate in the covalent $\sigma$-bonding with three neighboring P atoms, whereas the remaining two electrons occupy a lone pair orbital. Thus, for phosphorene and its derivatives such as P$_4$X (X=O and S), we considered van der Waals interaction through the non-local correlation functional optB88-vdW during the structural optimization,~\citep{PhysRevLett.92.246401,PhysRevLett.103.096102} while for the larger systems in few-layer and heterostructure phosphorene we used the D3 functional with zero damping.~\citep{1.3382344} The obtained lattice parameters for SLP are $a$ = 4.58 and $b$ = 3.32 \AA\ along the armchair and zigzag  directions, respectively, are consistent with the previous reports and experimental black phosphorus (Supplemental Material).~\citep{10.1038/ncomms5475,nl502892t,nl5032293,Brown:a04860,10.1063/1.438523, supple} 

The optimized structures with the PBE exchange-correlation functional are then used for the subsequent $GW$-BSE calculations. The quasiparticle (QP) picture is investigated within the partially self-consistent $GW_0$ approach by iterating the one-electron energies in the Green's function $G$.~\citep{PhysRev.139.A796, PhysRevB.75.235102} Two self-consistent updates for the Green's function ($G_2W_0$) are found to be sufficient to converge the QP band gap. The convergence of QP gap as a function of unoccupied bands was found to be much faster for phosphorene,~\citep{PhysRevLett.115.066403} and we find that 158 such bands to be sufficient in this regard. Further, the electron-hole interactions are incorporated within the BSE formalism, which provides the optical gap and the corresponding exciton binding energy.~\citep{RevModPhys.74.601,PhysRevLett.80.4510,PhysRevLett.81.2312}

The optical properties are calculated using frequency dependent complex dielectric tensor, $\varepsilon (\omega) = \varepsilon{'} (\omega) + i \varepsilon{''} (\omega)$. The imaginary part $\varepsilon{''} (\omega)$ of the linear dielectric tensor is calculated in the long-wavelength $\mathbf{q} \rightarrow \mathbf{0}$ limit,~\citep{PhysRevB.73.045112}
\begin{eqnarray}
\varepsilon''_{\alpha \beta}(\omega) & = & \frac{4\pi^2e^2}{\Omega} \lim\limits_{q\rightarrow 0}\frac{1}{q^2}\sum\limits_{c,v,\mathbf{k}}2w_{\mathbf{k}} \delta(\epsilon_{c\mathbf{k}} - \epsilon_{v\mathbf{k}} - \omega) \nonumber \\
&\times& \langle u_{c\mathbf{k}+\mathbf{e}_{\alpha}q} | u_{v\mathbf{k}} \rangle \langle u_{c\mathbf{k}+\mathbf{e}_{\beta}q} | u_{v\mathbf{k}} \rangle^*, 
\end{eqnarray}
where $\Omega$ is the volume of the primitive cell, $\omega_{\mathbf k}$ are $k$-point weights, and the factor 2 inside the summation accounts for the spin degeneracy. The $\epsilon_{c\mathbf{k}}$ ($\epsilon_{v\mathbf{k}}$) are ${\mathbf k}$-dependent conduction (valence) band energies, $u_{c\mathbf{k}, v\mathbf{k}}$ are cell periodic part of the pseudo-wave-function, and $\mathbf{e}_{\alpha, \beta}$ are unit vectors  along the Cartesian directions. The real part $\varepsilon{'} (\omega)$ is calculated using Kramers-Kronig transformation, and  the absorption co-efficient is calculated as $\Lambda_{\alpha \alpha} (\omega) = \frac{2\omega}{c}[|\varepsilon_{\alpha \alpha} (\omega)| - \varepsilon{'}_{\alpha \alpha} (\omega)]^{\frac{1}{2}}$. 

Within the BSE formalism, an exciton state $| S \rangle$ can be written as,~\citep{PhysRevLett.81.2312} 
\begin{equation}
| S \rangle = \sum\limits_{\mathbf{k}}\sum\limits_{v}^{\rm hole}\sum\limits_{c}^{\rm elec} A_{vc{\mathbf k}}^S  | vc{\mathbf k} \rangle ; 
\end{equation}
where $| vc{\mathbf k} \rangle = \hat{a}_{v{\mathbf k}}^{\dagger} \hat{b}_{c{\mathbf k+{\mathbf q}}}^{\dagger} | 0 \rangle$ with $| 0 \rangle$ being the ground state, and $\hat{a}^{\dagger} (\hat{b}^{\dagger})$ is hole (electron) creation operator. ${\mathbf q}$  is the momentum of the absorbed photon, and $A_{vc{\mathbf k}}^S$ are electron-hole amplitudes. The corresponding excitation energies $E_S$ are determined via BSE,~\citep{PhysRevB.29.5718}
\begin{eqnarray}
\left( \epsilon^{\rm QP}_{c{\mathbf k+{\mathbf q}}}  - \epsilon^{\rm QP}_{v{\mathbf k}}  \right)A_{vc{\mathbf k}}^S &+&  \sum\limits_{v'c'\mathbf{k'}} A_{v'c'{\mathbf k'}}^S \langle vc{\mathbf k} | K^{eh} | v'c'{\mathbf k'}   \rangle                         \nonumber \\
&=& E_S A_{vc{\mathbf k}}^S,  
\end{eqnarray}
where $ \epsilon^{\rm QP}$ are quasiparticle energies, and $K^{eh}$ is the electron-hole interaction. The imaginary part of the dielectric function $\varepsilon''(\omega)$ is calculated by the optical transition matrix element of the excitations. 

Such a rigorous treatment within the many-body theory coupled with BSE scheme provides an excellent description of QP and optical gaps; and exciton binding, which all compare well with the experimental results.~\citep{RevModPhys.74.601,PhysRevLett.111.216805,nmat4061,PhysRevB.87.155304,PhysRevB.86.115409} However, such treatment is restricted to the systems with small size due to its exceptional computational cost. Thus, one needs to develop simplified models to investigate excitons in realistic systems with appreciable accuracy. For three-dimensional materials, the simplistic Mott-Wannier model predicts the exciton binding energy as, $E_{\rm x}^{\rm 3D} = (\mu/\varepsilon^2)R_{\infty}$, where $R_{\infty}$ is the Rydberg constant. The excitonic effective mass $\mu$ and the static dielectric constant $\varepsilon$ can easily be calculated within the standard electronic structure calculations.~\citep{PhysRev.52.191} 

In contrast, the excitonic properties of two-dimensional materials are fundamentally different from their 3D counterpart, and cannot be described within the Mott-Wannier approach. In 2D materials, excitons are strongly confined and the dielectric screening is considerably reduced.~\citep{nphoton.2015.282,nl0716404, nl501133c, nl504868p, srep06608, nmat4061} There have been recent efforts to develop excitonic models for 2D materials, however, the significant effort has been devoted to the isotropic materials such as transition metal dichalcogenides.~\citep{PhysRevLett.116.056401,PhysRevB.88.045318, PhysRevLett.113.076802,PhysRevB.91.245421,PhysRevB.95.235434} Here, we present a generalized scheme appropriate for the anisotropic electronic materials such as phosphorene and its various derivatives.~\citep{PhysRevB.91.245421,PhysRevB.95.235434}  

For 2D semiconducting systems, the effective exciton Hamiltonian can be written as, 
\begin{equation}
H_{\rm x} = -\hbar^2 \frac{\nabla_r^2}{2\mu} + V_{\rm 2D}(r), 
\end{equation}
where $\mu^{-1} = m_e^{-1} +  m_h^{-1}$ is the exciton reduced mass, and $r$ is the electron-hole separation. Following Keldysh, the non-locally screened electron-hole interaction is described by,~\citep{Keldysh1979} 
\begin{equation}
V_{\rm 2D}(r) = - \frac{e^2}{4(\varepsilon_1+\varepsilon_2)\varepsilon_0r_0}\left[ H_0\left(\frac{r}{r_0}\right) - Y_0\left(\frac{r}{r_0}\right) \right], 
\end{equation}
where $\varepsilon_1$ and $\varepsilon_2$ are the dielectric constant of the upper and lower media; and $\varepsilon_0$ is the vacuum permittivity. $H_0$ and $Y_0$ are Struve and Bessel functions. The screening length $r_0$ is related to the 2D polarizability $\chi_{_{\rm 2D}}$ as $r_0 = 2\pi\chi_{_{\rm 2D}}$,~\citep{PhysRevB.84.085406} where $\chi_{_{\rm 2D}}$ is calculated using the static dielectric constant $\varepsilon$ of the concerned 2D material, $\varepsilon(L_{\rm v}) = 1 + 4\pi\chi_{_{\rm 2D}}/L_{\rm v}$, where $L_{\rm v}$ is the transverse vaccuum size. The $\varepsilon$ is calculated from the real part of complex dielectric tensor $\varepsilon(\omega)$ at zero frequency. Note that the interaction $V_{\rm 2D}$ at large separation $r>>r_0$ follow the $1/r$ Coulomb interaction, whereas at the $r<<r_0$ limit, the interaction reduces to a weaker $\log(r)$ dependence. 

The variational excitonic wave function, for an anisotropic electronic material such as phosphorene with $m_e^x \neq m_e^y$ and $m_h^x \neq m_h^y$, is written as, 
\begin{equation}
\psi(x,y) = 2\sqrt{\frac{2}{\pi \lambda a_x^2}}\exp\left[-\left\{ (x/a_x)^2 + (y/a_y)^2 \right\} ^{\frac{1}{2}} \right],
\end{equation}
where $a_y = \lambda a_x$ is the anisotropic exciton extension along the $x$ (armchair) and $y$ (zigzag) directions, respectively, and treated as variational parameters. Using this form of excitonic wave function $\psi(x,y)$, the expectation value of the kinetic energy is calculated to be, 
\begin{equation}
E_{\rm k}(\lambda, a_x) = \frac{\hbar^2}{4a_x^2}\left[ \frac{1}{\mu_x} + \frac{1}{\lambda^2\mu_y} \right].
\end{equation}
Here $\mu_x$ and $\mu_y$ are the reduced exciton mass along the $x$ and $y$ directions, respectively. The corresponding potential energy is given by,~\citep{PhysRevB.91.245421} 
\begin{equation}
E_{\rm p}(\lambda, a_x) = \int \int V_{\rm 2D}(x,y) |\psi(x,y)|^2 dxdy. 
\end{equation}
The variational exciton binding energy, $E_{\rm x}^{\rm 2D}(\lambda, a_x) = E_{\rm k}(\lambda, a_x) + E_{\rm p}(\lambda, a_x)$, is minimized with respect to the variational parameters $\lambda$ and $a_x$ to obtain the exciton binding energy, and (anisotropic) exciton extension. The parameters in the above model, effective electron and hole masses in different crystallographic directions and the static dielectric constant are then calculated from the first-principles calculations.  In principle, these parameters can be calculated using any level of approximation to the exchange-correlation functional, and here we have used the Heyd-Scuseria-Ernzerhof (HSE06) hybrid functional.~\citep{10.1063/1.1564060,10.1063/1.2404663}

\section{Results and Discussion}
First, we discuss the optical and excitonic properties of pristine phosphorene and investigate the effect of uniaxial strain along the different crystallographic directions. Here, we compare the computationally expensive $GW$-BSE results for these systems with the simplified model calculations. Once we demonstrate an excellent agreement between these methods, we extend our investigation within the simplified model for the realistically large systems, where $GW$-BSE calculations are practically impossible. 

\begin{figure}[t]
\begin{center}
\rotatebox{0}{\includegraphics[width=0.48\textwidth]{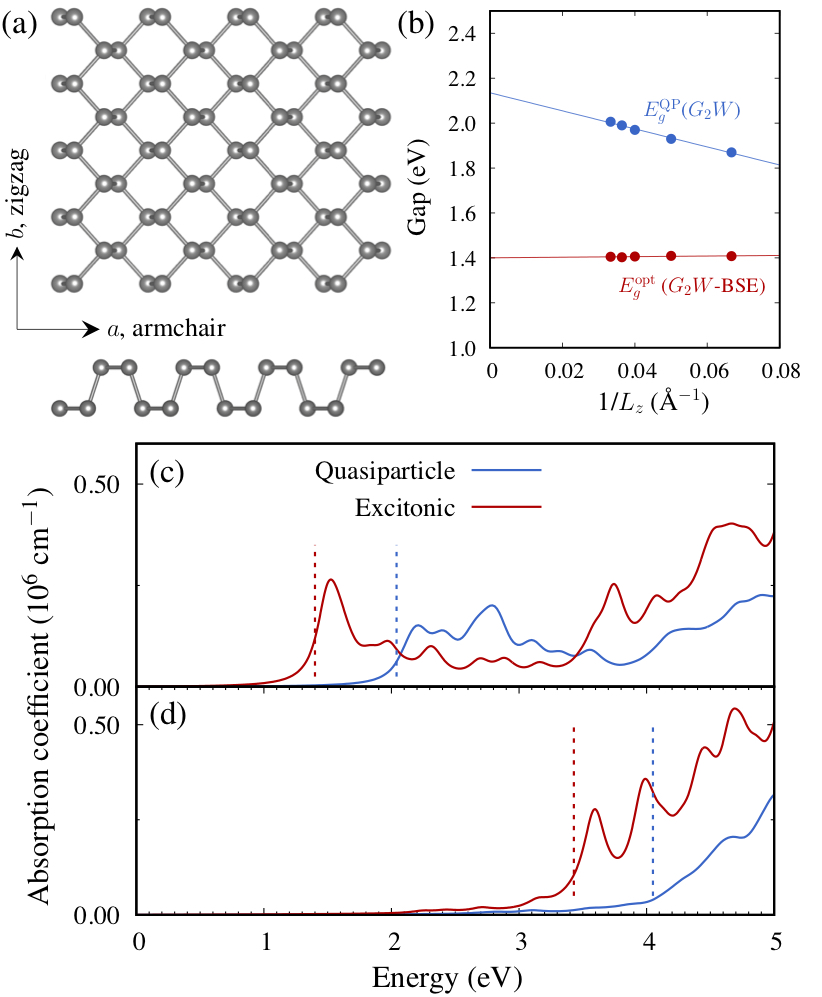}}
\caption{(a) The top and side view of single-layer phosphorene is shown indicating the armchair and zigzag crystallographic directions. (b) The QP and optical gaps are calculated with varied vertical separation $L_z$ between the layers and extrapolated to $L_z \rightarrow \infty$. The results are in excellent agreement with the experimental results.~\citep{nn501226z,nnano.2015.71,nl502892t} Absorption coefficient calculated without and with the electron-hole interaction for $L_z$=30 \AA. $\Lambda (\omega)$ shows strong absorption anisotropy between the light polarization along the (c) armchair and (d) zigzag directions. The vertical lines indicate the band edges corresponding to the first absorption peak. The difference between the QP and optical gaps indicates strongly bound exciton.}
\label{fig:figure1}
\end{center}
\end{figure}

\subsection{Pristine phosphorene} 
We start with the electronic and optical properties of single-layer pristine phosphorene [Figure~\ref{fig:figure1}(a)] within the $GW$-BSE approach. Due to the long-range Coulomb interactions, these properties of 2D materials are strongly influenced by the vertical separation $L_z$ between the periodic images.~\citep{PhysRevLett.115.066403,PhysRevLett.96.126104}  As it was shown earlier that the QP band gap converges as $1/L_z$, we extrapolate the gap to $L_z \rightarrow \infty$ limit, which we found to be 2.14 eV [Figure~\ref{fig:figure1}(b) and Table~\ref{tab:table1}].  This extrapolated $E_g^{\rm QP}$ is in excellent agreement with the previous theoretical,~\citep{PhysRevLett.115.066403} and with the two experimental results that are available to date.~\citep{nl502892t,nnano.2015.71}  The high-resolution transmission spectroscopy predicted a transport gap of 2.05 eV,~\citep{nl502892t} while the photoluminescence excitation spectroscopy suggests a QP gap of 2.2 $\pm$ 0.1 eV.~\citep{nnano.2015.71}  Comparing the present $G_2W_0$ results with earlier $G_0W_0$ calculations,~\citep{PhysRevB.89.201408,PhysRevB.89.235319} we argue that the self-consistent correction to the self-energy ($\Sigma = iGW$), through the Green's function update, is essential to predict the QP gap correctly.  It is important to note here that the optical gap converges much faster than the QP gap with varying $L_z$ [Figure~\ref{fig:figure1}(b)]. 

The absorption coefficient $\Lambda(\omega)$ calculated without and with the electron-hole interaction shows strong anisotropy for the light polarizations along the armchair and zigzag directions [Figures~\ref{fig:figure1}(c)-(d)]. The $L_z \rightarrow \infty$ extrapolated optical gap of 1.40 eV [Figure~\ref{fig:figure1}(b) and Table~\ref{tab:table1}] along the armchair direction  is in excellent agreement with the experimental range of 1.30 -- 1.45 eV.~\citep{nn501226z,nnano.2015.71} The large exciton binding energy of 0.74 eV indicates strongly bound exciton in SLP, which is in excellent agreement with the previous $GW$-BSE calculations~\citep{PhysRevLett.115.066403, PhysRevB.89.235319} and the experimental prediction of 0.9 $\pm$ 0.12 eV.~\citep{nnano.2015.71}. Note the absorption edge corresponding to light absorption along the zigzag crystallographic direction lies at much higher energy, and thus SLP shows linear dichroism.~\citep{10.1038/ncomms5475}

\begin{table*}[!t]
\caption{With varied $L_z$, the quasiparticle $E_g^{\rm QP}$ and optical $E_g^{\rm opt}$ gaps are calculated within the $G_2W_0$ and $G_2W_0$-BSE approaches, respectively, which are subsequently interpolated to $L_z \rightarrow \infty$. The parameters for the simplified exciton model, the effective carrier masses $m_e^*$ and $m_h^*$ along the armchair and zigzag directions; and the average static dielectric constant are calculated within the HSE06 hybrid functional. The corresponding 2D polarizability $\chi_{\rm 2D}$ is also tabulated. The uniaxial strain severely affects the many-body interaction in phosphorene, and thus the $E_g^{\rm QP}$, $E_g^{\rm opt}$, and $E_{\rm x}$ are altered. The effect of high impurity coverage is also investigated. The exciton binding energies $E_{\rm x}$ calculated within the hydrogenic exciton model compares excellently with those from the accurate and computationally expensive $GW$-BSE formalism, and available experimental results.}
\begin{tabular}{L{2.4cm} C{1.9cm} C{1.2cm} C{1.1cm} C{1.1cm}  C{1.3cm} C{1.3cm} C{1.3cm} C{1.3cm} C{1.1cm} C{1.1cm}}
\hline
\hline\\ [-0.2cm]
System & $E_g^{\rm QP}$ & $E_g^{\rm opt}$ & \multicolumn{2}{c}{$E_{\rm x}$ (eV)}& \multicolumn{2}{c}{armchair ($x$)} & \multicolumn{2}{c}{zigzag ($y$)} & $\varepsilon$ & $\chi_{\rm 2D}$ \\
              &       (eV)              &   (eV)             &  BSE  & model & $m_e^*/m_e$ & $m_h^*/m_e$ & $m_e^*/m_e$ & $m_h^*/m_e$ & & (\AA) \\ [0.1cm]
\hline \\ [-0.2cm]
SLP      & 2.05~\citep{nl502892t}    &  1.45~\citep{nn501226z}  &    &  & & & & & &\\
(experiment)     &   2.2 $\pm$ 0.1~\citep{nnano.2015.71}   & 1.30~\citep{nnano.2015.71} &  \multicolumn{2}{c}{0.9 $\pm$ 0.12~\citep{nnano.2015.71}} & & & & & & \\
SLP ($L_z \rightarrow \infty$) & 2.14 & 1.40 & 0.74 & 0.79 & 0.16 & 0.12 & 1.40 & 4.69 & 2.60 & 3.55 \\
SLP, $\varepsilon_s^a=$$-$5\% & 1.78 & 1.03 & 0.75 & 0.72 & 0.14 & 0.12 & 1.35 & 3.10 & 2.78 & 3.95 \\
SLP, $\varepsilon_s^a=$$+$5\% & 2.38 & 1.51 & 0.87 & 0.80 & 0.13 & 0.12 & 1.45 & 3.28 & 2.51 & 3.36\\
SLP, $\varepsilon_s^z=$$-$5\% & 1.89 & 1.12 & 0.77 & 0.78 & 0.23 & 0.13 & 1.39 & 2.64 & 2.70 & 3.77\\
SLP, $\varepsilon_s^z=$$+$5\% & 2.20 & 1.41 & 0.79 & 0.78 & 0.20 & 0.14 & 1.39 & 2.76 & 2.71 & 3.79\\
P$_4$O                        & 3.22 & 2.31 & 0.91 & 0.81 & 0.76 & 0.24 & 0.17 & 3.42 & 2.48 & 3.17\\
P$_4$S                        & 2.74 & 1.87 & 0.87 & 0.67 & 0.63 & 0.24 & 0.80 & 0.40 & 3.19 & 4.63\\
[0.1cm]
\hline
\hline\\
\end{tabular}
\label{tab:table1}
\end{table*}

We calculate the exciton binding within the hydrogenic effective exciton mass model, where the parameters are calculated from a relatively less computationally expensive treatment of the electron exchange and correlation, namely within the HSE06 hybrid functional.   The effective carrier masses are highly anisotropic (Table~\ref{tab:table1}) as estimated by fitting the HSE06 bands to the parabolic dispersion $E(k) = \hbar^2k^2/2m^*$. The estimated $m_e^*$ is much smaller 0.16$m_e$ along the armchair direction than the same along the zigzag direction with 1.40$m_e$, where $m_e$ is the electron rest mass. The qualitative picture is the same for the hole,  as $m_h^*$ along the armchair direction is much lighter 0.12$m_e$ compared to 4.69$m_e$ along the zigzag direction. Thus, the effective exciton mass along the armchair direction is much lighter compared to the same along the zigzag direction ($\mu_x$ $<<$ $\mu_y$). These results are in good agreement with the previous results.~\citep{10.1038/ncomms5475,PhysRevB.90.085402}   The other parameter of the model, static dielectric constant is calculated from the real part of the complex dielectric function at zero frequency. The average of static $\varepsilon$  along the armchair and zigzag directions is used to calculate the 2D polarizability $\chi_{\rm 2D}$ (Table~\ref{tab:table1}). Using these parameters, the effective exciton mass model predicts an exciton binding energy of 0.79 eV, which is in excellent agreement with the present $GW$-BSE prediction and the experimental estimation (Table~\ref{tab:table1}). 

The spatial distribution of exciton state is found to be anisotropic and extended along the armchair direction ($a_x$=12.26 and $a_y$=4.90 \AA), with spatial anisotropy satisfying the relation, $\lambda = a_y/a_x \sim (\mu_x/\mu_y)^{1/3}$, that was analytically predicted earlier.~\citep{PhysRevB.91.245421}  In an agreement, such elliptic spatial structure of bound hole has recently been observed in black phosphorus through STM tomographic imaging.~\citep{nanolett.7b03356}

\subsection{Effect of uniaxial strain}
\begin{figure*}[!t]
\begin{center}
\rotatebox{0}{\includegraphics[width=0.96\textwidth]{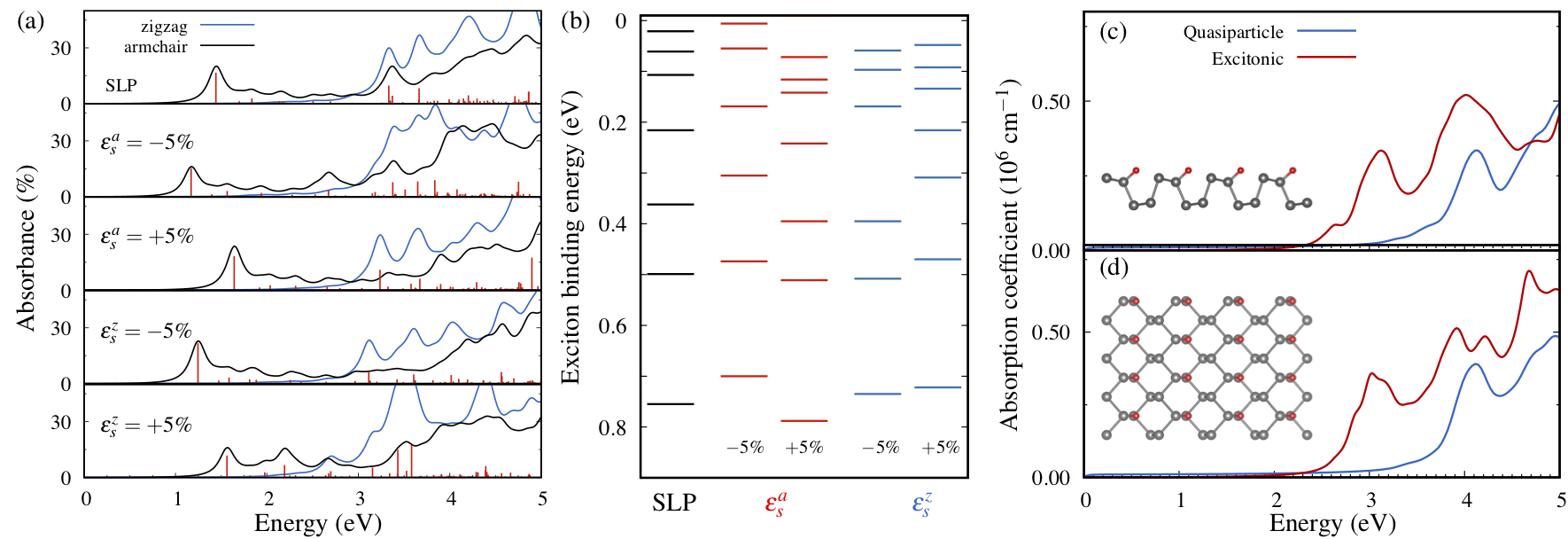}}
\caption{(a) The effects of strain on the absorbance and dipole oscillator strength calculated including the electron-hole interaction within the $GW$-BSE formalism ($L_z$ = 30 \AA). The vertical lines indicate the relative dipole oscillator strengths and are significantly affected by the applied uniaxial strain. (b) The  corresponding exciton spectra for the pristine and uniaxially strained SLP.  Strain-dependent dielectric screening influences the spectra. The $E=0$ eV refers to the quasiparticle gap, and excitons generated from the transitions with non-vanishing dipole oscillator strengths are shown. The relative oscillator strengths are shown in (a). Absorption coefficient calculated without and with the electron-hole interaction ($L_z$ = 30 \AA) for the light polarization along (c) armchair and (d) zigzag directions, shown for 1 ML O coverage. The absorption edges are calculated from the corresponding $E$ vs $(E\Lambda)^{1/2}$ plot for this indirect gap semiconductor. The absorption anisotropy along the armchair and zigzag directions is greatly reduced due to monolayer impurity coverage. The insets show the side and top views of SLP with 1ML O coverage.}
\label{fig:figure2}
\end{center}
\end{figure*}

The effect of uniaxial strain on the electronic structure of SLP has been studied earlier within the conventional exchange-correlation functional.~\citep{PhysRevB.90.085402,PhysRevB.90.205421,10.1021/jp508618t, PhysRevB.97.045132} However, the considerations of self-energy correction and electron-hole interaction are scarce in this regard. Here, we investigate the QP and optical gaps; and the corresponding exciton binding for SLP under $\varepsilon^{a/z}_s=\pm 5$\% uniaxial strain (Table~\ref{tab:table1}), while the strained lattice is relaxed along the transverse direction. We earlier reported that the strain energy along the zigzag direction is much higher compared to the same along the armchair direction and thus, straining the SLP along the armchair direction is comparatively easier.~\citep{PhysRevB.97.045132}  Within $-$5\% $\leqslant$ $\varepsilon^{a/z}_s$ $\leqslant$ $+$5\% strain range, the $L_z \rightarrow \infty$ interpolated gaps $E_g^{\rm QP}$ and $E_g^{\rm opt}$ decrease with uniaxial compressive strain, while they increase with tensile strain (Table~\ref{tab:table1}).   Clearly, the many-body interaction is comparatively more affected by the uniaxial strain along the armchair direction. Further, the compressive strain triggers a stronger renormalization than the tensile strain of equal magnitude. The absorption anisotropy in different crystallographic directions remains intact [Figure~\ref{fig:figure2}(a)]. Within the applied strain range, the first absorption peak appears within 1$-$1.50 eV for light polarization along the armchair direction. In contrast, for the light polarization along the zigzag direction, the first peak appears above 3 eV and in general shows higher absorbance.  Applied uniaxial strain changes the oscillator strength significantly [Figure~\ref{fig:figure2}(a)]. While the oscillator strength corresponding to the ground state exciton changes by $\pm$10 \% for the strain $\varepsilon_s^a=\pm 5\%$ along the armchair direction, the change is even more significant, $\mp$30\% for applied strain along the zigzag direction, $\varepsilon_s^z=\pm 5\%$. Interestingly, for $\varepsilon_s^z =$5\%, we observe a competing absorption peak just below its QP gap. Such strain dependent $E_g^{\rm QP}$ and $E_g^{\rm opt}$ may explain the variation observed in photoluminescence peak on different substrates.~\citep{nn501226z,nnano.2015.71} 

The excitons remain strongly bound under applied strain, while the $G_2W_0$-BSE calculated $E_{\rm x}$ varies between 700--900 meV  within the investigated strain range. The subsequent excited states reflect a sensitive dependence on the strain dependent dielectric screening [Figure~\ref{fig:figure2}(b)]. Thus, strain engineering in phosphorene leads to wide photoluminescence energy, enhanced absorption, and  multiple exciton formation.

Next, we investigate the effective carrier masses with a varied uniaxial strain within the HSE06 functional. In agreement with an earlier report,~\citep{PhysRevB.90.085402} the effective electron and hole masses are severely affected by strain (Table~\ref{tab:table1}), however their qualitative anisotropic nature remains intact with $\mu_x << \mu_y$.  These $m_e^*$ and $m_h^*$  are used to estimate $E_{\rm x}$ for the strained SLP within the hydrogenic model, which are in excellent agreement with the more accurate $GW$-BSE results (Table~\ref{tab:table1}). Further, the spatial anisotropy of exciton $\lambda \sim (\mu_x/\mu_y)^{1/3}$ remain similar under strain as in pristine SLP (Supplemental Material).~\citep{supple}  These results essentially validate the applicability of the present hydrogenic model for the anisotropic 2D phosphorene.

\subsection{Effect of impurity coverage}
The presence of lone-pair electrons in phosphorene makes it reactive and can easily absorb impurities with a strong binding energy resulting from the P to impurity change transfer. We have earlier discussed O, S and N chemisorption with varied impurity coverage and it was concluded that 0.25--0.5 monolayer (ML) O/S coverages become conducive to both water redox reactions.~\cite{10.1021/acs.jpcc.7b12649}  While the exciton binding energy plays an important role in efficient charge separation and in turn affects the performance of a catalytic device, the SLP derivatives with such low impurity coverages are very difficult to investigate within the $GW$-BSE formalism. Thus at first, we investigate the derivatives with high 1 ML O/S coverages, P$_4$O and P$_4$S that can be represented by a small cell, and compare the $GW$-BSE results with the effective mass model. 

The SLPs with 1 ML O and S coverage are found to be indirect gap semiconductors, and both QP and optical gaps are severely altered (Table~\ref{tab:table1}). While for the 1 ML O-covered SLP the extrapolated $E_g^{\rm QP}$ and $E_g^{\rm opt}$ increase upon impurity coverage, the picture is reversed for the 1 ML S-covered SLP.  The absorption edge, calculated by considering the electron-hole interaction, lies at a much lower energy than the same calculated without the interaction (Figure~\ref{fig:figure2}). This observation indicates strongly bound excitons in these derivatives similar to the pristine SLP (Table~\ref{tab:table1}).  The qualitative anisotropic feature in the carrier effective masses are intact but are severely altered in all crystallographic directions.~\citep{supple} However, the anisotropy in effective exciton mass disappears with $\lambda \sim 1$, and thus the corresponding exciton extension becomes isotropic with 1 ML coverage (Supplemental Material).~\citep{supple}  
Furthermore, the $E_{\rm x}$ calculated within the $GW$-BSE formalism are in excellent agreement with those calculated using the hydrogenic model (Table~\ref{tab:table1}), which implies the applicability of this simplistic model to investigate exciton in such phosphorene derivatives with impurity coverage.
    
The SLPs with sub-monolayer coverages are both thermodynamically and kinetically stable. Further, the band edges align with the redox potentials for water splitting reactions for SLPs with 0.33$-$0.5 ML oxygen/sulfur coverages.~\citep{10.1021/acs.jpcc.7b12649} Carrier effective masses in all directions are severely affected by the impurity coverage, and the resulting exciton binding energies are found to be high (Supplemental Material).~\citep{supple}  The calculated $E_{\rm x}$ within the hydrogenic model is found to be 0.81 and 0.85 eV (0.85 and 0.86 eV) for 0.33 and 0.5ML oxygen (sulfur) coverages. Such high exciton binding makes charge separation difficult in optoelectronic and catalytic devices.  Thus, although the conduction and valence bands in these derivatives align with the redox potentials, the catalytic activity is expected to be negatively impacted due to high exciton binding. On the other hand, the high absorption coefficient [Figure~\ref{fig:figure2}] and robust exciton with high binding energy are desirable features for light emitting device applications. In this regard, the impurity covered phosphorene can extend such application to green-light emission.\citep{nnano.2014.26,nmat4205} The anisotropy in exciton extension is in agreement with the analytical estimation of $\lambda \sim (\mu_x/\mu_y)^{1/3}$, which monotonically decreases with increasing coverage.~\citep{supple}

\subsection{Exciton renormalization in few-layer and heterostructure phosphorene}
\begin{figure}[!t]
\begin{center}
\rotatebox{0}{\includegraphics[width=0.49\textwidth]{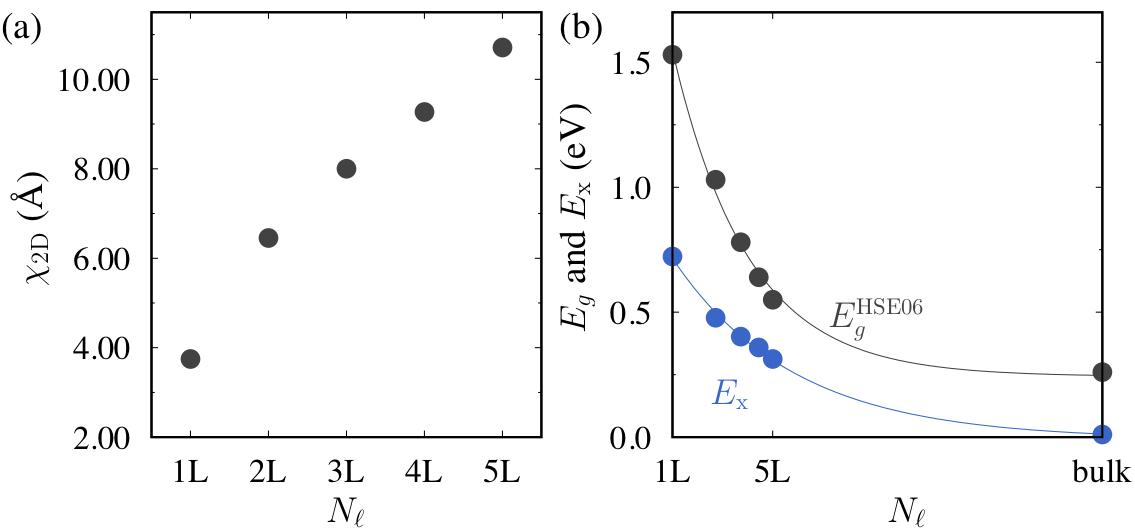}}
\caption{(a) Calculated $\chi_{\rm 2D} = (\varepsilon - 1)L_{\rm v}/4\pi$ increases with layer thickness $N_{\ell}$ indicating an increase in electron screening. (b) The renormalization of bandgap $E_g$ and exciton binding $E_{\rm x}$ in few-layer phosphorene show a power-law dependence with the layer thickness and both quantities converge very slowly to the corresponding bulk value. The variation in exciton binding is largely determined by the effective hole mass $m^*_h$ along the zigzag direction, in addition to the change in effective electron screening.}
\label{fig:figure3}
\end{center}
\end{figure}

While the layer-dependent evolution of gaps has been studied in few-layer phosphorene,~\citep{10.1038/ncomms5475, PhysRevB.89.235319,10.1021/nn503893j,nnano.2016.171} it is intriguing to investigate the layer-dependent exciton binding.  The effective screening increases with layer thickness and in addition, the carrier masses along different crystallographic directions are also expected to be modified.  The evolution in lattice parameters with increasing layer thickness agrees well with the previous prediction and converges to the bulk values (Supplemental Material).~\citep{supple,10.1038/ncomms5475}  Further, the bandgap in few-layer phosphorene decreases with thickness and converges to the value for black phosphorus [Figure~\ref{fig:figure3}(b)]. The calculated bulk gap of 0.26 eV agrees reasonably well with the experimental measurement of 0.33 eV.~\citep{JPSJ.52.2148}  Considering the fact that the gaps are underestimated within the HSE06 functional, it is imperative to investigate its qualitative dependence on thickness.  A power-law fit $E_g = aN_{\ell}^{-\alpha} + c$, with $N_{\ell}$ being the number of layers, indicates that $E_g$ decays much slower as $\alpha = 0.83$ than the usual quantum confinement with $\alpha = 2$.  While $\alpha < 2$ is generic in weak van der Waals stacked two-dimensional materials, calculated $\alpha$ in few-layer phosphorene is much smaller compared to MoS$_2$, where $\alpha = 1.10$.~\citep{PhysRevLett.105.136805} 

The effective electron screening increases with thickness $N_{\ell}$, and the static dielectric constant $\varepsilon$ in few-layer phosphorene increases with the number of layers. The anisotropic $\varepsilon$ in black phosphorus ($\varepsilon_x$ = 14.75, $\varepsilon_y$ = 10.87, $\varepsilon_z$ = 8.68) are consistent with the experimental measurement of 16.5, 13 and 8.3 along the armchair, zigzag and perpendicular directions, respectively.~\citep{JPSJ.54.2096} The dependence of $\chi_{\rm 2D}$, calculated using the average of anisotropic static $\varepsilon$, indicates increasing screening with thickness [Figure~\ref{fig:figure3}(a)].   

Effective carrier masses are mostly unaffected in few-layer phosphorene except for the hole mass along the zigzag direction, which exhibits a strong layer dependence (Supplemental Material).~\citep{supple}  The calculated $m^*_h$ along this direction decreases sharply with thickness $N_{\ell}$, which is in agreement with a previous prediction.~\citep{10.1038/ncomms5475} To confirm the evolution of effective carrier masses with $N_{\ell}$, we calculated the same for the bulk black phosphorus, which compares well with the experimental values (Supplemental Material).~\citep{supple, 978-1-4615-7682-2_299}

Consequently, the renormalization of exciton binding in few-layer phosphorene [Figure~\ref{fig:figure3}(a)] is dictated by the strong dependence of effective hole mass along the zigzag direction and the increase in effective screening with layer thickness. This result is in contrast to the assumption that the exciton binding and thus its variation is independent of carrier effective mass.~\citep{PhysRevLett.116.056401, Zhangeaap9977} The large exciton binding sharply decreases from 0.72 eV in 1L to 0.48 eV in 2L and 0.31 eV in 5L phosphorene, which is still much higher than the corresponding bulk value. We estimated the Mott-Wannier exciton binding in black phosphorus $E_{\rm x}^{\rm 3D} = (\mu/\varepsilon^2)R_{\infty}$ to be about 11 meV, which is much smaller than $k_BT$ at room temperature, and also much smaller than the same in bulk transition-metal dichalcogenides.~\citep{APL1.4945047}  A similar power-law fit with $\alpha=0.53$ indicates a much slower dependence of $E_{\rm x}$ on thickness than for $E_g$. Further, the exciton extension in few-layer phosphorene remains anisotropic while the degree of anisotropy decreases monotonically with thickness as $\lambda$ increases (Supplemental Material).~\citep{supple}

\begin{figure}[!t]
\begin{center}
\rotatebox{0}{\includegraphics[width=0.48\textwidth]{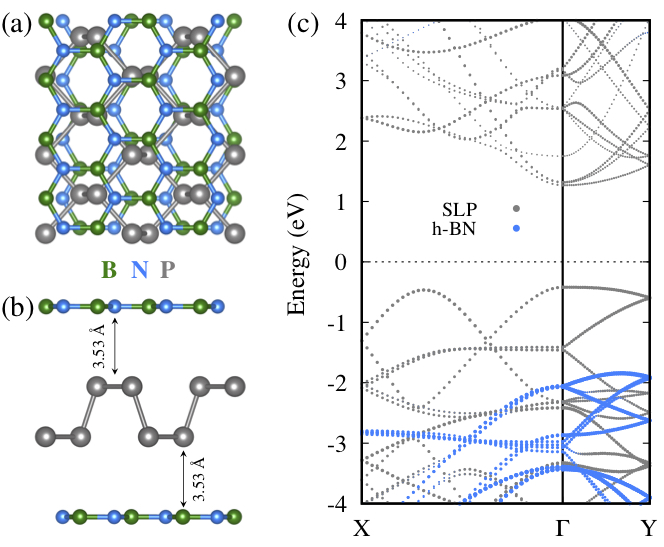}}
\caption{(a)-(b) The top and side view of the h-BN/SLP/h-BN van der Waals encapsule. The structure is optimized using the PBE+D3-vdW functional with zero damping. The encapsulation induces a compressive (tensile) strain of 5.2\% (1.3\%) in the phosphorene layer along the armchair (zigzag) direction. (c) The valence band maxima and conduction band minima originate from the SLP and thus indicate a type I band alignment. While the apparent band structure looks similar to the pristine SLP, the corresponding carrier effective masses are severely altered, which in turn modifies the exciton character. }
\label{fig:figure4}
\end{center}
\end{figure}

Black phosphorous is often encapsulated to avoid degradation and restore the intrinsic electronic properties.~\citep{nnano.2015.71,nl5032293,10.1038/nnano.2014.35} The hexagonal boron nitride (h-BN) was demonstrated to be a good candidate material for this purpose.~\citep{acsnano.5b00289,ncomms7647,ncomms8315} Thus, it would be worth investigating that how the electronic structure and effective dielectric screening are affected due to the substrate and capping. In this regard, we investigate the h-BN/SLP and h-BN/SLP/h-BN heterostructures [Figure~\ref{fig:figure4}(a) and (b)]. To minimize the lattice strain in the phosphorene layer, we used a 1$\times$3 supercell of phosphorene with an orthorhombic 1$\times$4 cell of h-BN. Due to its puckered structure, the phosphorene lattice is relatively much softer than the h-BN and can easily sustain a large strain. Thus, in relaxed geometry, the strain is induced in the phosphorene layer while the h-BN lattice is mostly unaltered.~\citep{supple} The bandgap of SLP in these heterostructures is only slightly modified compared to the pristine case (Supplemental Material).~\citep{supple} We attribute this small change in the gap to the induced strain in the phosphorene layer in the heterostructure. These results are consistent with the previous calculations.~\citep{acs.nanolett.6b00154,2053-1583-5-4-045031}   
 
Previously, the exciton binding energy was proposed to be independent of the effective mass.~\citep{PhysRevLett.116.056401} In contrast, within the present model, $E_{\rm x}$ is determined by the effective mass and 2D polarizability. Moreover, the spatial anisotropic structure is directly related to the carrier effective masses, $\lambda \sim (\mu_x/\mu_y)^{1/3}$. Thus, in addition to the dielectric environment, any change in the intrinsic electronic structure in phosphorene due to the substrate and capping layer is important to consider for a correct description of the exciton. Indeed, the effective carrier mass in the phosphorene layer is affected by h-BN,~\citep{supple} which along with the electron screening from the h-BN layer reduces the exciton binding. The exciton binding is renormalized to 0.55 eV for the SLP/h-BN heterostructure, while the presence of a second h-BN layer in the h-BN/SLP/h-BN encapsulation does not alter the exciton binding further (0.53 eV). The excitons in these heterostructures are generated in the phosphorene layer, and no interlayer exciton is possible [Figure~\ref{fig:figure4}(c)].  Further, the spatial anisotropy of excitons in these heterostructures is reduced.~\citep{supple} Similar renormalization of exciton binding was predicted earlier for Al$_2$O$_3$/phosphorene/h-BN encapsulation, and phosphorene on SiO$_2$ or PDMS substrates.~\citep{nanolett.7b01365, 2053-1583-1-2-025001, Zhangeaap9977} 


\section{Summary}
We have studied the quasiparticle and optical properties of phosphorene and investigated the role of uniaxial strain, and impurity coverages.  The excitonic properties described within a tractable anisotropic hydrogenic model exhibit excellent agreements with those calculated by explicitly considering the electron-hole interaction within the $GW$-BSE formalism.  In contrast to the previous assumption, the exciton binding strongly depends on effective carrier masses, which further determines the anisotropic spatial extension of excitons. Similar to the pristine SLP, exciton in strained and impurity covered phosphorene remain strongly bound. However, due to the change in the corresponding carrier masses, the anisotropy in spatial extension is severely altered. Owing to a severe alteration in the effective hole mass along the zigzag direction and increase in the electron screening, the exciton binding is greatly renormalized in few-layer phosphorene. In contrast, the exciton binding is relatively less affected in the h-BN and phosphorene heterostructures. The robust large exciton binding energy and tunable photoluminescence in encapsulated and impurity covered phosphorene derivatives raise their prospective applications in light-emitting devices. The results indicate that the present model will be applicable to other phosphorene based superstructures and other two-dimensional anisotropic materials.

\begin{acknowledgements}
We acknowledge the supercomputing facilities at the Centre for Development of Advanced Computing, Pune; Inter University Accelerator Centre, Delhi; and at the Center for Computational Materials Science, Institute of Materials Research, Tohoku University. M.K. acknowledges funding from the Department of Science and Technology through Nano Mission project SR/NM/TP-13/2016 and the Science and Engineering Research Board for Ramanujan Fellowship.
\end{acknowledgements}


%

\clearpage


\section*{Supplementary material}

# Supplemental Material for
# Exciton in phosphorene: Strain, impurity coverage, thickness and encapsulation

Srilatha Arra,[1, *] Rohit Babar,[2, *] and Mukul Kabir[2, 3, †]

[1] *Department of Chemistry, Indian Institute of Science Education and Research, Pune 411008, India*
[2] *Department of Physics, Indian Institute of Science Education and Research, Pune 411008, India*
[3] *Centre for Energy Science, Indian Institute of Science Education and Research, Pune 411008, India*

(Dated: October 29, 2018)

* These authors contributed equally

† Corresponding author: mukul.kabir@iiserpune.ac.in

TABLE S1. Evolution of lattice parameters with layer thickness in few-layer phosphorene. Results are in good agreement with the experimental lattice parameters of bulk black phosphorus (BP). Lattice parameters slightly differ within different treatment of van der Waals interaction. For the vdW-D3 correction we used the zero damping method. The optimized structures with optB88 and D3 vdW corrections have been used to calculate the HSE06 bandstructure.

|  |  | Lattice parameter (Å) | | HSE06 gap (eV) |
|---|---|---|---|---|
|  |  | $a$ (armchair) | $b$ (zigzag) |  |
| 1L | vdW-optB88 | 4.58 | 3.32 | 1.54 |
|  | vdW-D3 | 4.59 | 3.30 | 1.53 |
| 2L | vdW-D3 | 4.51 | 3.30 | 1.03 |
| 3L | vdW-D3 | 4.49 | 3.30 | 0.78 |
| 4L | vdW-D3 | 4.47 | 3.30 | 0.64 |
| 5L | vdW-D3 | 4.45 | 3.30 | 0.51 |
| Bulk BP | vdW-D3 | $a$=4.42, $b$=3.31, $c$=10.67 | | 0.26 |
|  | vdW-optB88 | $a$=4.48, $b$=3.34, $c$=10.72 | | 0.26 |
|  | Experiment [1, 2] | $a$=4.37, $b$=3.31, $c$=10.47 | | 0.33 [3] |

TABLE S2. The effect of uniaxial strain along the armchair and zigzag directions in pristine SLP. The bandgap decreases (increases) with compressive (tensile) strain in the studied strain range of $\varepsilon_{a/z} \leqslant 5\%$. The anisotropy in the exciton extension $\lambda$ is in excellent agreement with the analytical prediction of $(\mu_x/\mu_y)^{1/3}$, and largely remains unaltered under strain.

| System | HSE06 Gap(eV) | $\lambda$ | $(\mu_x/\mu_y)^{1/3}$ | $a_x$ (Å) | $a_y$ (Å) |
|---|---|---|---|---|---|
| SLP | 1.54 | 0.40 | 0.40 | 12.26 | 4.90 |
| SLP, $\varepsilon_a$ =−5% | 1.30 | 0.41 | 0.41 | 13.20 | 5.40 |
| SLP, $\varepsilon_a$ =+5% | 1.73 | 0.39 | 0.39 | 12.59 | 4.91 |
| SLP, $\varepsilon_z$ =−5% | 1.39 | 0.45 | 0.45 | 11.62 | 5.19 |
| SLP, $\varepsilon_z$ =+5% | 1.65 | 0.44 | 0.44 | 11.65 | 5.17 |

TABLE S3. Effective carrier masses for SLP with O/S impurity coverages and $\chi_{2D}$ calculated within HSE06 functional. These parameters are used with the hydrogenic model to calculate the exciton binding energy $E_x$ and the corresponding exciton extension anisotropy $\lambda$.

| System | armchair | | zigzag | | $\varepsilon$ (avg) | $\chi_{2D}$ (Å) | $E_x$ (eV) | $\lambda$ |
|---|---|---|---|---|---|---|---|---|
|  | $m_e^*/m_e$ | $m_h^*/m_e$ | $m_e^*/m_e$ | $m_h^*/m_e$ |  |  |  |  |
| 0.33 ML O | 0.31 | 0.16 | 0.43 | 3.15 | 2.56 | 3.33 | 0.81 | 0.65 |
| 0.50 ML O | 0.19 | 1.48 | 0.46 | 3.71 | 2.66 | 3.54 | 0.85 | 0.74 |
| 1 ML O | 0.76 | 0.24 | 0.17 | 3.42 | 2.48 | 3.17 | 0.81 | 1.04 |
| 0.33 ML S | 0.60 | 0.16 | 0.94 | 2.61 | 2.74 | 3.62 | 0.85 | 0.56 |
| 0.50 ML S | 0.33 | 1.16 | 0.52 | 1.27 | 2.82 | 3.81 | 0.86 | 0.88 |
| 1 ML S | 0.63 | 0.24 | 0.80 | 0.40 | 3.19 | 4.63 | 0.67 | 0.86 |

TABLE S4. Effective carrier masses and $\chi_{2D}$ for few-layer phosphorene calculated within HSE06 functional. These parameters are used with the hydrogenic model to calculate the exciton binding energy $E_x$ and the corresponding exciton extension anisotropy $\lambda$. Exciton binding for the bulk BP is calculated within the Mott-Wannier model.

| System | vdW | armchair | | zigzag | | $\varepsilon$ (avg) | $\chi_{2D}$ (Å) | $E_x$ (eV) | $\lambda$ |
|---|---|---|---|---|---|---|---|---|---|
|  |  | $m_e^*/m_e$ | $m_h^*/m_e$ | $m_e^*/m_e$ | $m_h^*/m_e$ |  |  |  |  |
| 1L | optB88 | 0.16 | 0.12 | 1.40 | 4.69 | 2.60 | 3.55 | 0.793 | 0.40 |
| 1L | D3 | 0.12 | 0.10 | 1.38 | 5.34 | 2.69 | 3.75 | 0.723 | 0.37 |
| 2L | D3 | 0.12 | 0.11 | 1.45 | 1.95 | 4.59 | 6.45 | 0.478 | 0.42 |
| 3L | D3 | 0.12 | 0.11 | 1.43 | 1.76 | 5.98 | 8.00 | 0.402 | 0.46 |
| 4L | D3 | 0.13 | 0.12 | 1.42 | 1.25 | 6.86 | 9.27 | 0.359 | 0.49 |
| 5L | D3 | 0.13 | 0.12 | 1.41 | 1.03 | 7.51 | 10.71 | 0.313 | 0.57 |
| BP | D3 | 0.12 | 0.10 | 1.39 | 0.82 | 11.43 | − | 0.011 | − |

TABLE S5. Calculated carrier effective masses calculated for bulk BP within the HSE06 hybrid functional. Present results compare well with an earlier theoretical calculation and experimental estimation.

|  |  | $a$, armchair | | $b$, zigzag | | $c$, perpendicular | |
|---|---|---|---|---|---|---|---|
|  |  | $m_e^*/m_e$ | $m_h^*/m_e$ | $m_e^*/m_e$ | $m_h^*/m_e$ | $m_e^*/m_e$ | $m_h^*/m_e$ |
| Bulk BP | Present | 0.12 | 0.10 | 1.39 | 0.82 | 0.17 | 0.30 |
|  | Ref. [4] | 0.12 | 0.11 | 1.15 | 0.71 | 0.15 | 0.30 |
|  | Experiment [5] | 0.082 | 0.078 | 1.027 | 0.648 | 0.128 | 0.28 |

TABLE S6. Lattice parameter and the corresponding HSE06 gap for the h-BN/SLP heterostructure and the h-BN/SLP/h-BN encapsulate. The gap is mostly unaffected, while the slight change in gap is attributed to the strain induced in the SLP layer in the heterostructures. Note that, while the SLP layer in both the heterostructures have similar lattice parameters, they produce different bandgaps. This is largely due to the change in the distance between the two-half layers of SLP (Figure S1).

| System | Lattice parameters (Å) | | Strain in phosphorene layer | | Bandgap (eV) |
|---|---|---|---|---|---|
|  | $a$ (armchair) | $b$ (zigzag) | $\varepsilon_a$ | $\varepsilon_z$ |  |
| Pristine SLP | 4.59 | 9.90 | — | — | 1.53 |
| Pristine h-BN | 4.35 | 10.05 | — | — | 5.71 |
| SLP/h-BN | 4.36 | 10.02 | −5.01 | 1.21 | 1.45 |
| SLP in SLP/h-BN | 4.36 | 10.02 | −5.01 | 1.21 | 1.34 |
| h-BN/SLP/h-BN | 4.35 | 10.03 | −5.23 | 1.31 | 1.73 |
| SLP in h-BN/SLP/h-BN | 4.35 | 10.03 | −5.23 | 1.31 | 1.65 |

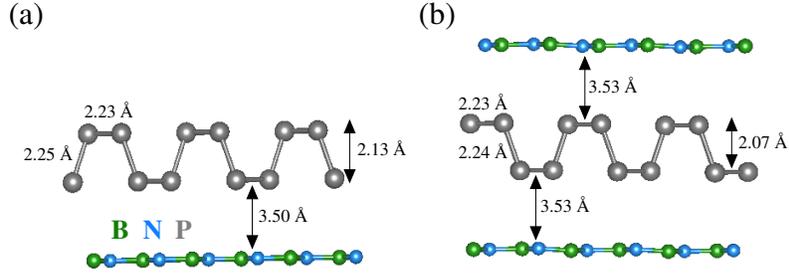

FIG. S1. (a) SLP/h-BN and (b) h-BN/SLP/h-BN heterostructures. The vertical separation between the layers changes slightly in these heterostructure. While the strain induced in the SLP layer for both these heterostructures are very similar, the distance between the two half-layers of phosphorene differs substantially by about 0.06 Å, which in turn is responsible for very different badgap.

TABLE S7. Carrier effective masses, $\chi_{2D}$, exciton binding $E_x$ and spatial anisotropy $\lambda$ for phoshorene heterostructures with h-BN.

| Heterostructure | $a$, armchair | | $b$, zigzag | | $\chi_{2D}$ (Å) | $E_x$ (eV) | $\lambda$ |
|---|---|---|---|---|---|---|---|
|  | $m_e^*/m_e$ | $m_h^*/m_e$ | $m_e^*/m_e$ | $m_h^*/m_e$ |  |  |  |
| SLP/h-BN | 0.26 | 0.23 | 0.68 | 4.51 | 3.71 | 0.55 | 0.59 |
| h-BN/SLP/h-BN | 0.23 | 0.32 | 0.28 | 4.60 | 3.44 | 0.53 | 0.81 |



\end{document}